%% file: eprint.tex
\documentclass[10pt]{article}
\usepackage{graphicx}

\def\Title#1{\begin{center} {\Large #1 } \end{center}}
\def\Author#1{\begin{center}{ \sc #1} \end{center}}
\def\Address#1{\begin{center}{ \it #1} \end{center}}

\newcommand\pubblock{\rightline{\begin{tabular}{l} Proceedings of the Fifth Annual LHCP\\ \pubnumber\\
         \pubdate  \end{tabular}}}

\newenvironment{Abstract}{\begin{quotation} \begin{center} 
             \large ABSTRACT \end{center}\bigskip 
      \begin{center}\begin{large}}{\end{large}\end{center} \end{quotation}}

\newenvironment{Presented}{\begin{quotation} \begin{center} 
             PRESENTED AT\end{center}\bigskip 
      \begin{center}\begin{large}}{\end{large}\end{center} \end{quotation}}

\def\Acknowledgements{\bigskip  \bigskip \begin{center} \begin{large}
             \bf ACKNOWLEDGEMENTS \end{large}\end{center}}

\input econfmacros.tex

\usepackage{color}

\newcommand\pubnumber{LHCb-PROC-2017-027}

\newcommand\pubdate{\today}

\def\affiliation{
On behalf of the LHCb Collaboration, \\
LAL\\ Universit\'e Paris-Sud,  CNRS$/$IN2P3, Orsay, France}

\begin{document}

\large
\begin{titlepage}
\pubblock

\vfill
\Title{Fixed-target physics at LHCb}
\vfill

\Author{\'Emilie Maurice}
\Address{\affiliation}
\vfill

\begin{Abstract}
The LHCb experiment has the unique possibility, among the LHC experiments, to be operated in fixed target mode, using its internal gas target SMOG. The energy scale achievable at the LHC and the excellent detector capabilities for vertexing, tracking and particle identification allow a wealth of measurements of great interest for cosmic ray and heavy ions physics. We report the first measurements made in this configuration:  the measurement of antiproton production in proton-helium collisions  and the measurements of open and hidden charm production in proton-argon collisions at  $\sqrt{s_\textrm{NN}} =$ 110 GeV.

\end{Abstract}
\vfill

\begin{Presented}
The Fifth Annual Conference\\
 on Large Hadron Collider Physics \\
Shanghai Jiao Tong University, Shanghai, China\\ 
May 15-20, 2017
\end{Presented}
\vfill
\end{titlepage}
\def\thefootnote{\fnsymbol{footnote}}
\setcounter{footnote}{0}
%

\normalsize 


\section{Introduction}

The LHCb experiment~\cite{LHCb} is a single-arm forward spectrometer covering the pseudorapidity range 2 $< \eta <$ 5. It was designed for the study of hadrons containing b or c quarks, and thus has excellent vertexing, tracking and particle identification (PID) capabilities. On top of that, LHCb provides the unique opportunity at the LHC to operate in a fixed target mode, thanks to the System for Measuring Overlap with Gas (SMOG). Originally designed for precise luminosity measurements~\cite{Lumi}, SMOG allows to inject noble gas such as Argon or Helium inside the
primary LHC vacuum around the LHCb vertex detector (VELO).
 Since 2015, LHCb has started to exploit SMOG to perform  physics
runs, using special fills not devoted to $pp$ physics, with different beam and target configurations,
allowing unique production studies which are relevant to cosmic ray and heavy ions physics. The first measurements using this unique configuration are reported in this document.

\section{$\overline{p}$ cross section measurement in $p$He collisions}


Recent results of the space-borne PAMELA and AMS-02 experiments on the $\overline{p}/p$ ratio  point to a higher yield of antiprotons and a milder energy dependence with respect to most theoretical prediction in the region above 50 GeV. 
The interpretation of these results, which provide a sensitive indirect probe for Dark Matter, is currently limited by the accuracy of the predicted secondary antiproton flux.
As there existed no data for $\overline{p}$ production in $p$He collisions in the 10-100 GeV energy range so far, 
the state-of-the-art calculations used extrapolations from the NA49 $pp$ and $p-$C data and other measurements, leading to an uncertainty of 20$-30 \%$, the single largest  contribution to the error on the $\overline{p}/p$ ratio prediction.

The production of antiprotons from the interaction of primary cosmic rays with the interstellar medium can be reproduced  by collisions of multi-TeV protons on helium at rest.
Thanks to its internal gas target, LHCb performed the first measurement
of antiproton production from collisions of 6.5 TeV LHC  proton beam on helium nuclei at rest. The measurement covers the antiprotons produced directly in the $p$He collisions, or from resonances decaying via the strong interaction, but not from the hyperon decays~\cite{Antiproton}.

\subsection{Selection and analysis}

The analyzed dataset corresponds to an integrated luminosity of 0.4 nb$^{-1}$, recorded in May 2016. 
Events were triggered with a minimum bias requirement, fully efficient on the events containing an antiproton within the LHCb acceptance. 
To benefit from the optimal reconstruction efficiency, the antiprotons candidates coming from collisions in a 80 cm long fiducial region along the beam direction $z$ are considered. 
The analysis is done in two-dimensionnal bins, in momentum $p$ and transverse momentum $p_T$, in the range $12 < p < 110$ GeV$/c$ and $0.4 < p_T < 4$ GeV$/c$. These limits are fixed by the acceptance of the two ring-imaging Cherenkov detectors.

The antiproton analysis requires a negative track associated to a primary vertex, both having quality requirements. 
The reconstruction and primary vertex efficiencies are evaluated using the simulation. 
Due to the large fiducial region, the primary vertex efficiency varies with $z$ from 76 $\%$ in the most upstream region to 95 $\%$ around the nominal LHCb collision point. 
The reconstruction efficiency including acceptance effects and the tracking efficiency is determined from simulation in three-dimensional bins of $p, p_T, z$. This tracking efficiency predicted by the simulation, averaged over $z$, ranges from 40 to 80 $\%$, depending on the track kinematics. To take into account the imperfections of the tracking detector response in the simulation, a correction of $\sim$ 1 $\%$ is calculated from calibration samples in $pp$ data.

The negative track is identified by two ring-imaging Cherenkov detectors. The fraction of antiprotons is extracted from the two-dimensional distribution describing the difference of the log likehood between the proton and pion, the proton and the kaon hypothesis. In each kinematic bin, the fraction of antiprotons is determined with a two-dimensional extended binned maximum likelihood fit whose templates are extracted from three set of data. The data set considered are: the calibration samples in the $p$He data, the calibration samples in the 2016 LHCb $pp$ data and the simulation. For each kinematic region, the most appropriate calibration templates are chosen, while the related systematic uncertainty is estimated from their comparison. 




For this analysis, only prompt antiprotons are considered. Therefore the nonprompt antiprotons are considered as background and removed from the sample by requiring an impact parameter compatible with zero. The residual nonprompt background is  due in 90$\%$ of cases to hyperon decays while the remaining 10 $\%$ correspond to antiprotons coming from a secondary collision in the detector material. The average background level is estimated to be $(2.6 \pm 0.6) \%$.
Another source of background is the possible contamination of the gas target, due to the LHC residual vacuum. To evaluate this additional source of background, a specific data taking was performed, using the same vacuum-pumping configuration but without injecting gas. The yield measured during this data taking is scaled according to the number of protons on target. The relative average contribution from residual gas vacuum is evaluated to be $(0.7 \pm 0.2) \%$.


The SMOG system does not presently allow a precise calibration of the injected gas pressure. Therefore the normalization for the $\overline{p}$ production is done by measuring a process with a well-known cross-section. Single electrons scattered off by the proton beam can be observed within the LHCb acceptance, and for a 6.5 TeV proton beam, in this kinematic range, the scattering is purely elastic. Only electrons in the polar angle $3 < \theta < 27$ mrad can be reconstructed in LHCb, leading to a cross-section of 180.6 $\mu$b. Thanks to a distinct signature, low $p$, $p_T$ track with little activity, these events can be identified even if they are three orders of magnitude below the total nuclear inelastic cross-section. The reconstruction efficiency, evaluated from simulation, is limited by the soft momentum and transverse momentum spectrum. 
Two kind of background are considered:  soft nuclear interaction  with the electron coming from a photon conversion and light hadron from a central exclusive production event. These backgrounds are charge symmetric while the signal is asymmetric. Therefore  the background can be modeled and subtracted. 
Possible contributions of gas ionization were evaluated to be negligible. 
Finally the luminosity is computed from the background-subtracted yield of scattered electrons, the known cross-section $\sigma_{pe^-}$ and the electron reconstruction efficiency. 
The measured luminosity is ${\cal{L}} = 0.443 \pm 0.011 \pm 0.027$ nb$^{-1}$, where the first uncertainty is statistical and the second systematic.

\begin{figure}
	\begin{center}
		\includegraphics[width=0.5\textwidth]{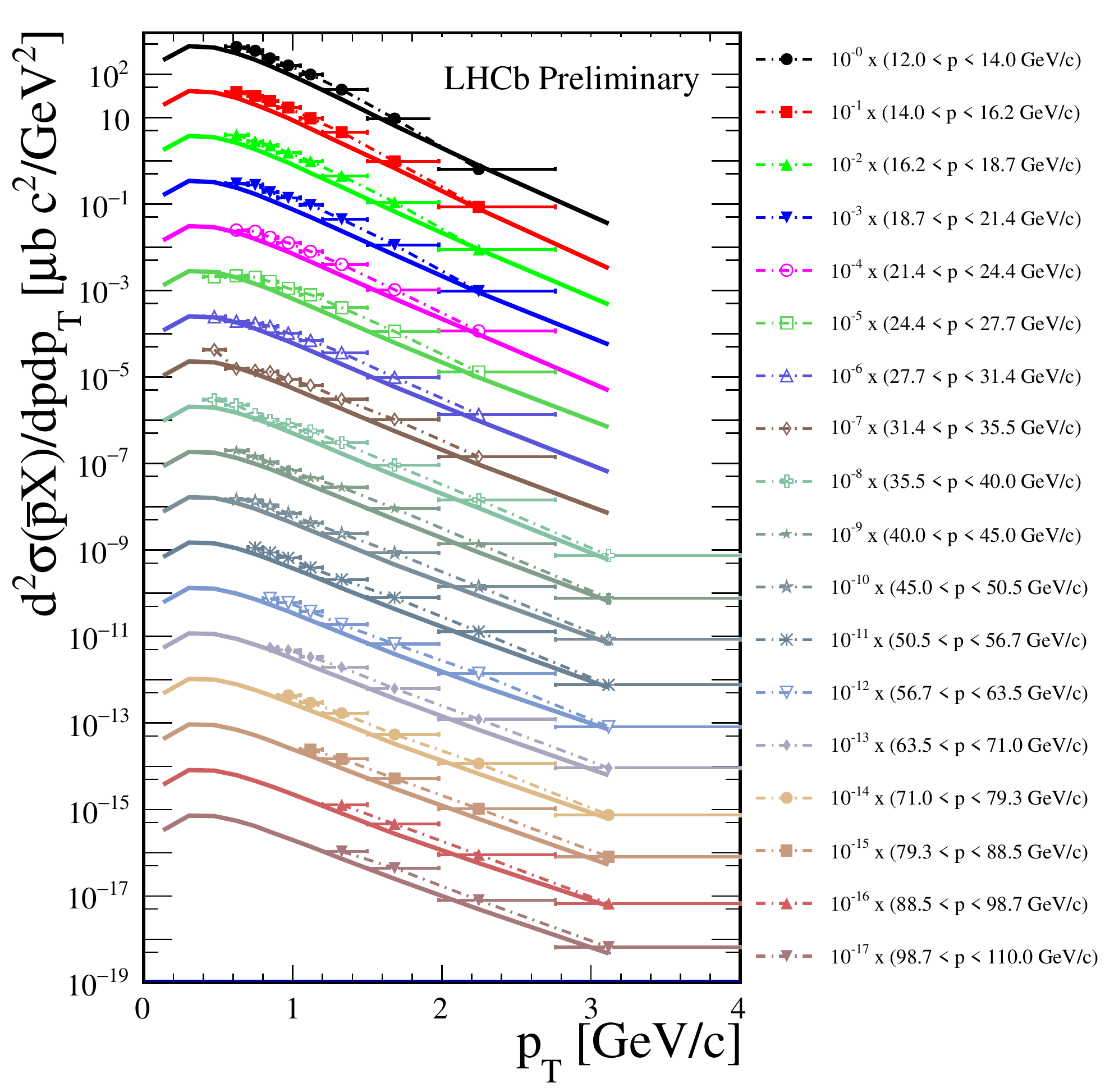}
			\caption{
			Result for the $\overline{p}$ cross-section measurement. The data points show the double differential cross-section as a function of $p_T$ in the 18 momentum bins, with values successively scaled by a factor 0.1 to improve the readibility of the plot. The solid curves show the EPOS LHC absolute predictions, scaled with the same factors as the data. The errors bars, barely visible, show the uncorrelated uncertainty only.
			\label{fig1_pHe}}
	\end{center}
\end{figure}

\subsection{Results}

Among 33.7 millions selected $p$He collisions, a sample of 1.4 million antiprotons was extracted. The double differential $\overline{p}$ production cross section is computed in each kinematic bin ($p$, $p_T$), after correcting for the reconstruction efficiency and the background. The precision of the measurements is limited by the systematic uncertainty. The largest correlated uncertainty among the bins  is due to the normalization, while the uncorrelated systematic uncertainties are dominated by the error on the $\overline{p}$ fraction from the particle identification analysis. At the borders of the detector acceptance and in the transition between the two RICH detector, the relative uncertainties are up to 26 $\%$. For the other regions, the uncertainty is typically a few per cent. For most of the accessible $p_T$ region, the relative total uncertainty amounts to 10 $\%$. 
The results presented on Fig.~\ref{fig1_pHe} are compared with the prediction of the EPOS LHC~\cite{EPOS} model, used in the LHCb simulation. The double differential shape is in good agreement with the simulation, while the absolute production rate is larger on average by about a factor 1.5. The data are also compared with other models implemented in the CRMC~\cite{CRMC} package v1.5.6: EPOS 1.99~\cite{EPOS_1.9.9}, HIJING 1.38~\cite{HIJING} and QGSJET II-0.4~\cite{QGSJET}.

The total inelastic cross-section is also determined from the measured total yield of recorded collisions within the fiducial region. The result is $\sigma_{\textrm{inel}}^{LHCb}(p\textrm{He}, \sqrt{s_{NN}} = 110 \textrm{GeV}) = (140 \pm 10) $ mb, which is larger than the EPOS LHC prediction by a factor 1.19 $\pm$ 0.08, implying that the measured $\overline{p}$ multiplicity per inelastic collisions is significantly larger in data.

\section{$J/\psi$ and $D^0$ production in $p$Ar data}

The LHCb fixed-target mode allows to study the heavy ions physics too. Production of heavy quarks in nucleus-nucleus interactions is a relevant probe of the transition between ordinary hadronic matter and the Quark-Gluon Plasma (QGP), in the high-density and high-temperature regime of QCD. 
As the heavy quarks are only produced in the early stage of the interaction, their total yields should not be modified by the QGP, while the relative yields of hidden heavy-flavour bound states such as charmonium are expected to be significantly modified~\cite{Satz_2006}.
Experimentally, the study of the sequential suppression of charmonium bound states is optimal at low energy, at a centre-of-mass energy  of $\sqrt{s_{NN}} \sim 100$ GeV, to avoid secondary charmonia production via statistical combination. 
The  interpretation of the charmonium suppression patterns observed in heavy-ion collisions needs the identification of phenomena not related to QGP 
which can be studied in  proton-induced reactions on  nuclear targets,    insensitive  to  effects  due  to  confinement. 
The first analysis of heavy-flavour production measured by the LHCb experiment in a fixed-target mode is presented:  $J/\psi$ and $D^0$ production processes in  proton-Argon sample  at $\sqrt{s_{NN}} = 110.4$ GeV~\cite{pArnote}.

\begin{figure}
	\begin{center}
		\includegraphics[width=.45\textwidth]{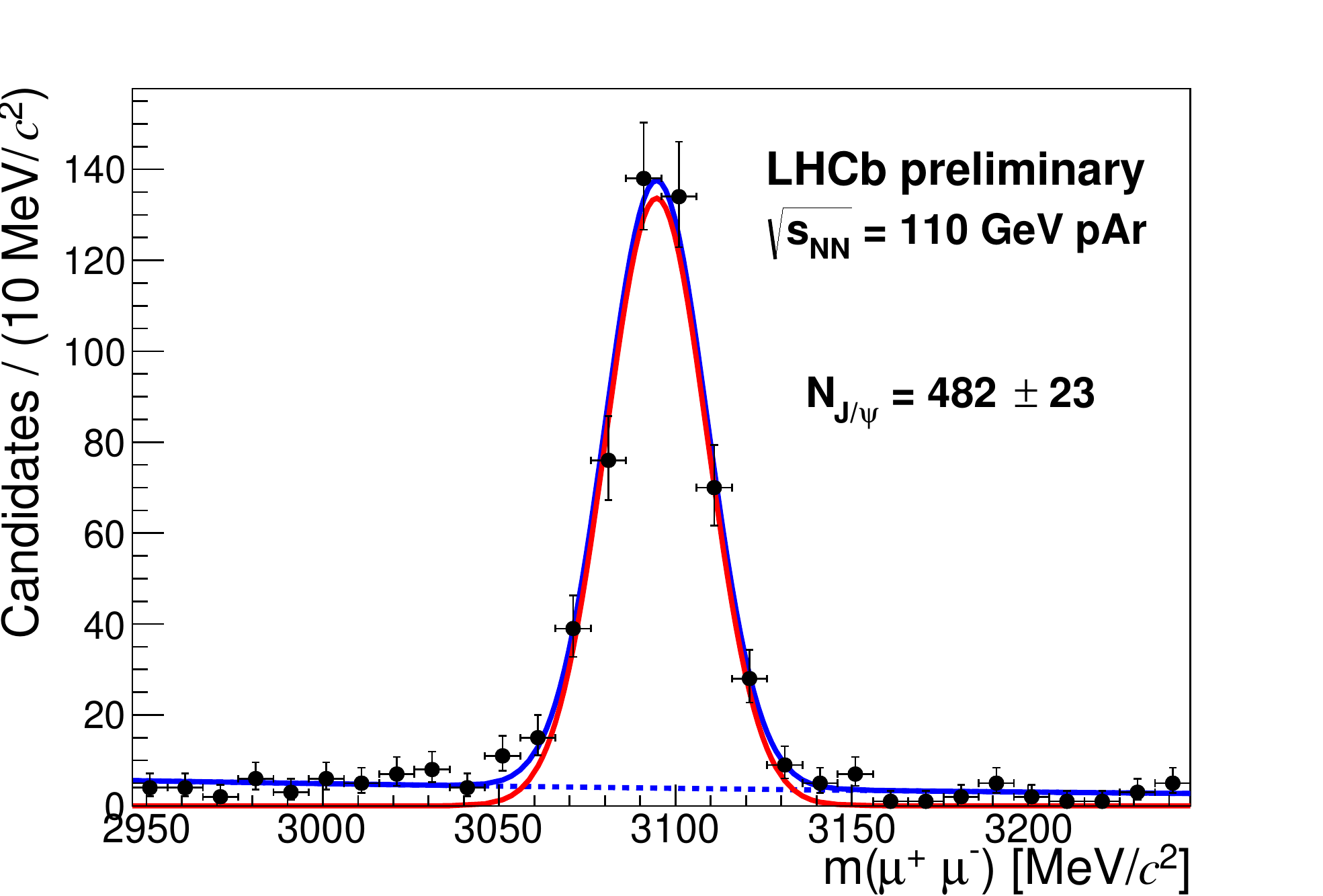}
		\includegraphics[width=.45\textwidth]{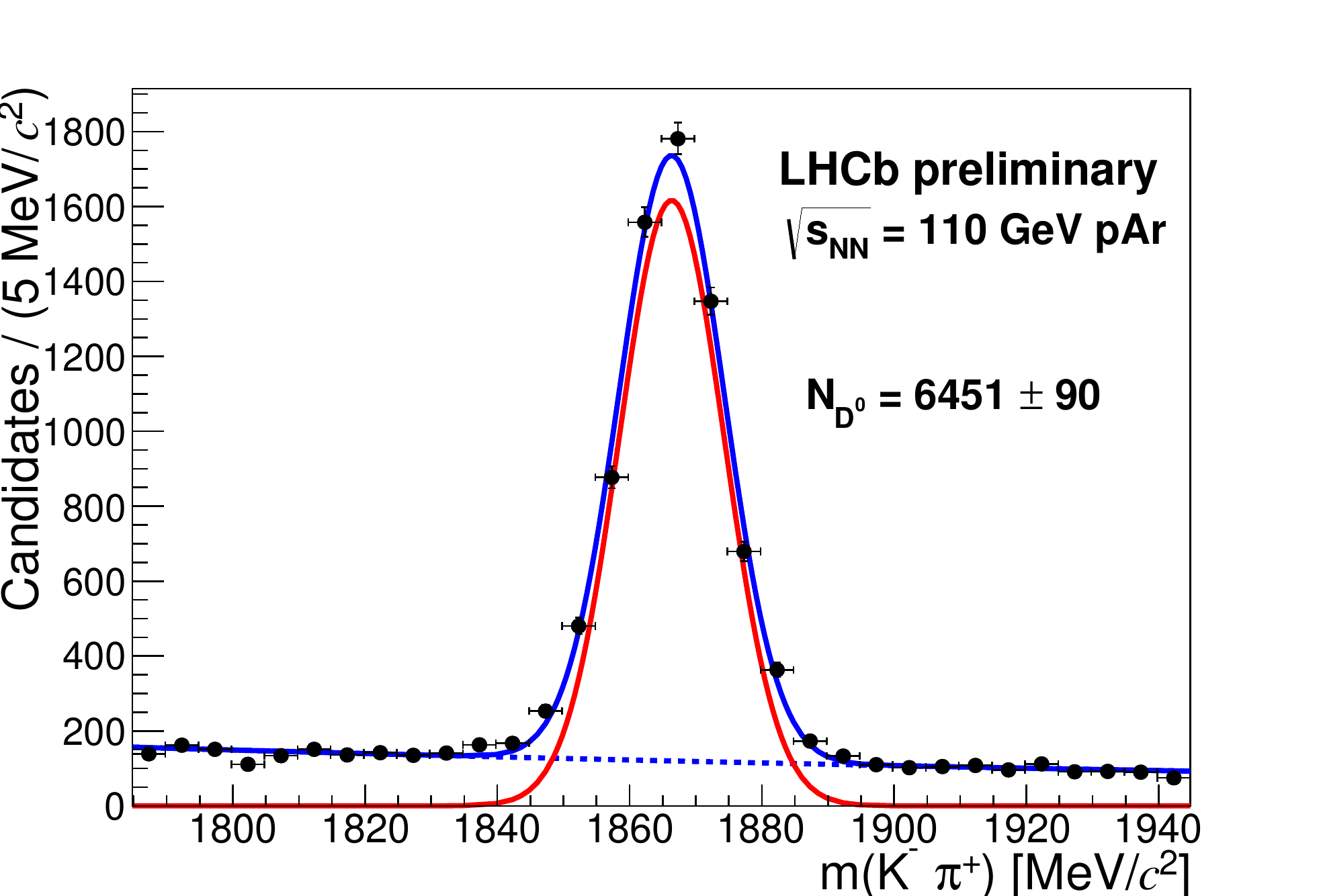}
		\caption{$J/\psi \to \mu^+ \mu^-$ (left) and $D^0 \to K^\mp \pi^\pm$ (right) mass distributions with the fit functions compared. The black points are the data, the red lin\
			e the signal, the dashed blue line the background, and the solid blue the sum of background and signal.
			\label{fig1}}
	\end{center}
\end{figure}

\subsection{Selection and analysis}
The study reported here  uses a data sample of $p$Ar collisions of a 6.5 TeV$/$c proton beam with the Argon gaseous target. 
The analysis studies the $J/\psi$ and $D^0$ candidates reconstructed in $J/\psi \to \mu^+ \mu^-$ and $D^0 \to K^\mp \pi^\pm$ decays. In  order  to  fully  benefit  from  the  LHCb detector's  performance, only events with a primary vertex reconstructed between -200 and 200 mm with respect to the nominal interaction point are considered.  
The numbers of $J/\psi$ and  $D^0$ signal candidates are extracted from unbinned maximum likelihood fits to their mass distributions, as shown in Fig~\ref{fig1} covering the full rapidity and transverse momentum ranges. 
For differential studies, the number of signal candidates is determined independently as a function of their transverse momentum $p_T$ and rapidity $y$. 

The yields determined from the mass fit are corrected for event trigger and selection efficiencies, including primary vertex and particle reconstruction, geometrical acceptance of the detector and particle identification.  Except for particle identification and tracking efficiencies that are obtained based on $pp$ data~\cite{LHCb-PUB-2016-021}, all the efficiencies are determined from fully simulated $p$Ar events.
 
 \begin{figure}
 	\begin{center}
 		\includegraphics[width=.4\textwidth]{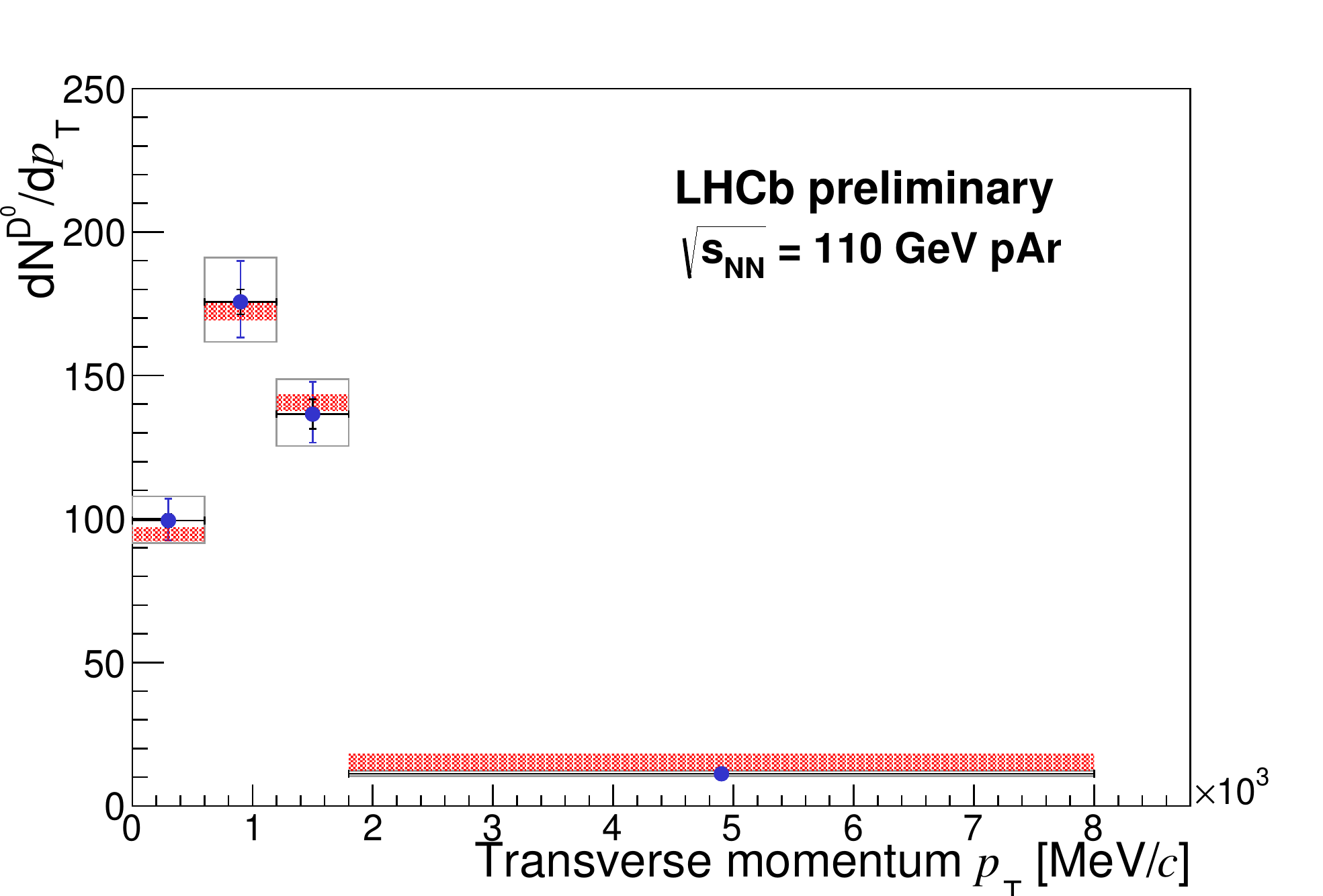}
 		\includegraphics[width=.4\textwidth]{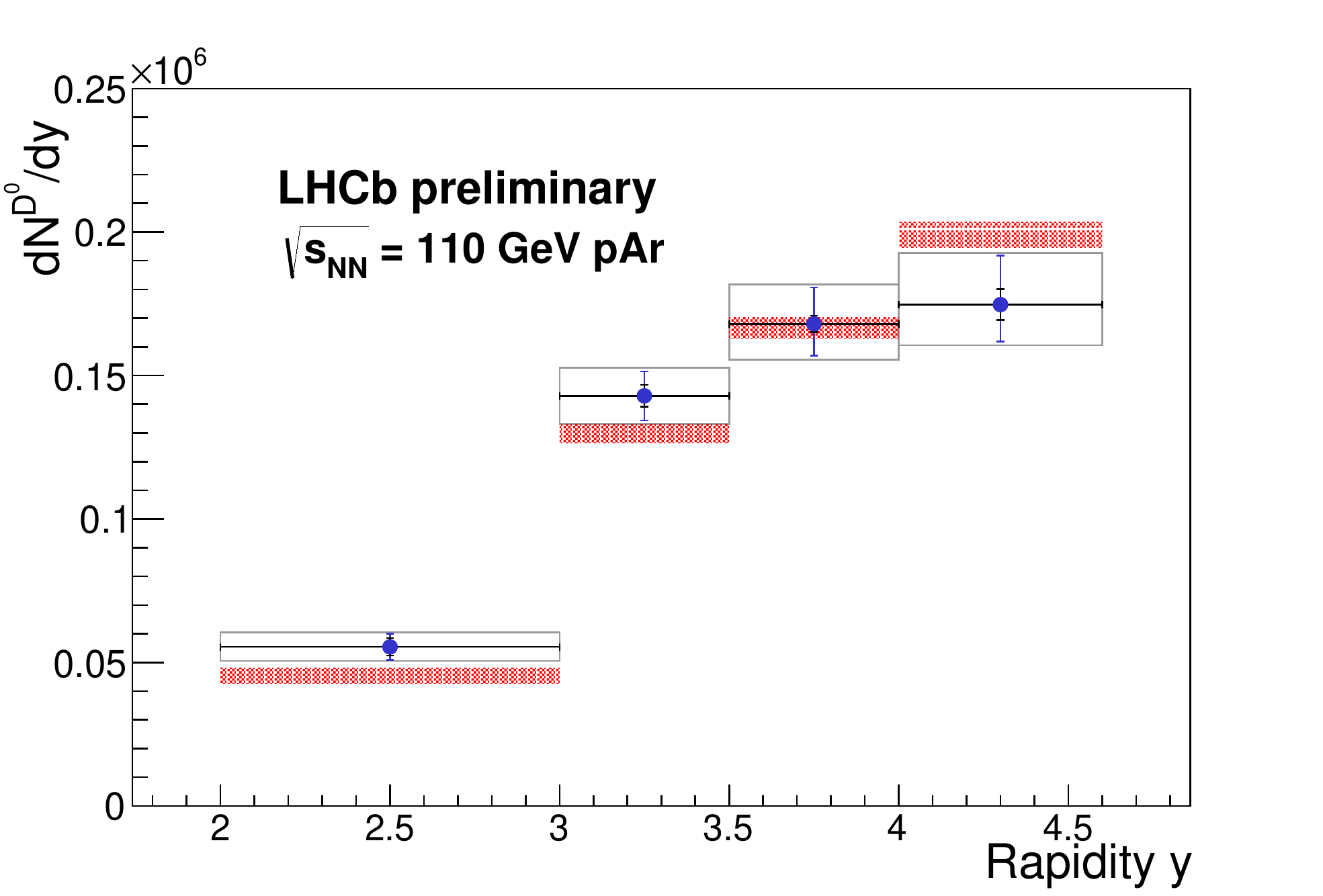}
 		\caption{$D^0$ differential corrected yields, as a function of transverse momentum $p_T$ (left) and rapidity $y$ (right). Red boxes correspond to  			Pythia predictions. Statistical uncertainties and the quadratic sum of statistical and uncorrelated systematic uncertainties are indicated by two different vertical lines  			(black line: statistical only; blue line: statistical+systematic). Overall systematic uncertainties, included the correlated systematic uncertainties, are quadratically added to the statistical and uncorrelated systematics and indicated by open boxes. \label{fig2}}
 	\end{center}
 \end{figure}

 \subsection{Results}
 \label{}
 
\begin{figure}
 	\begin{center}
 		\includegraphics[width=.4\textwidth]{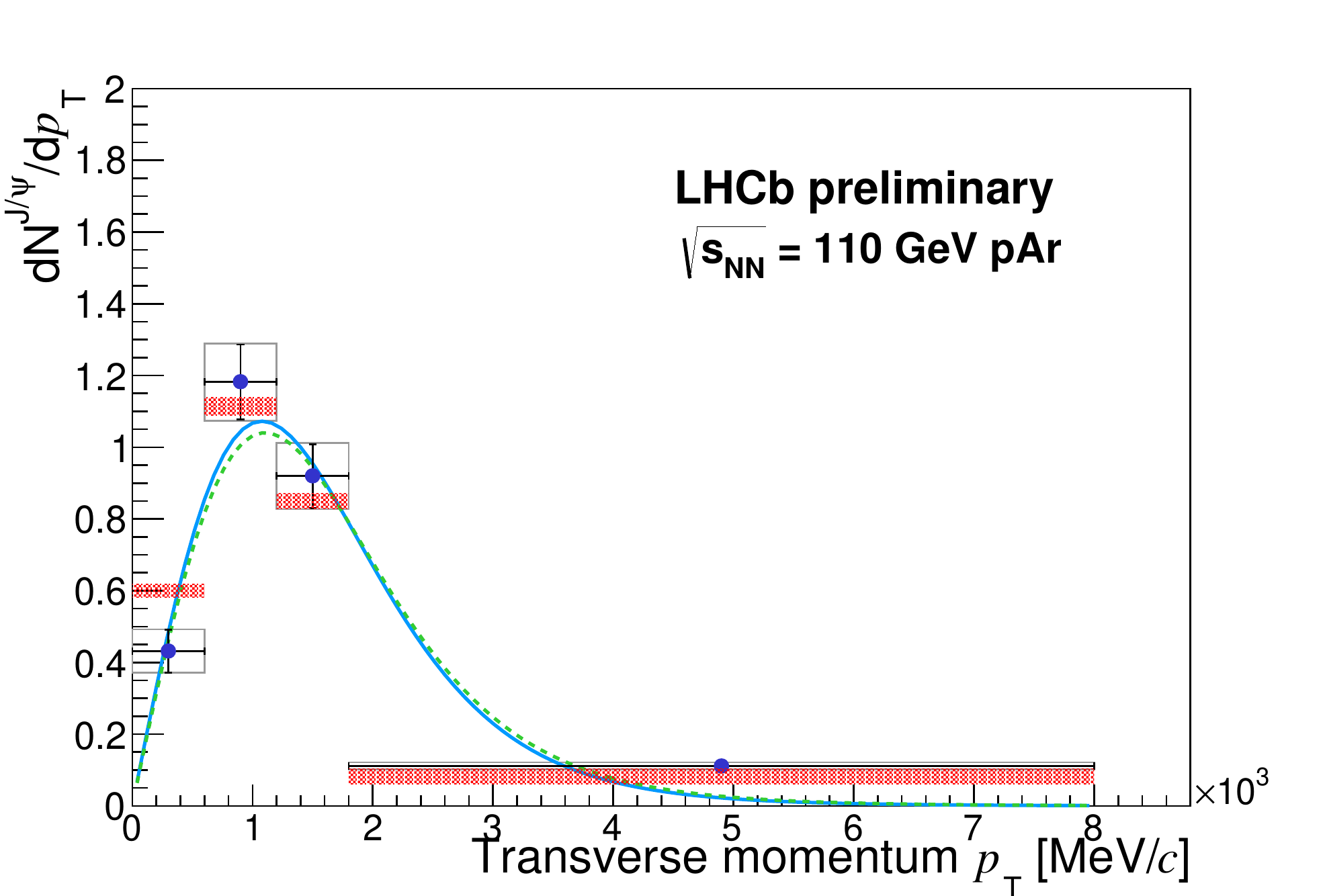}
 		\includegraphics[width=.4\textwidth]{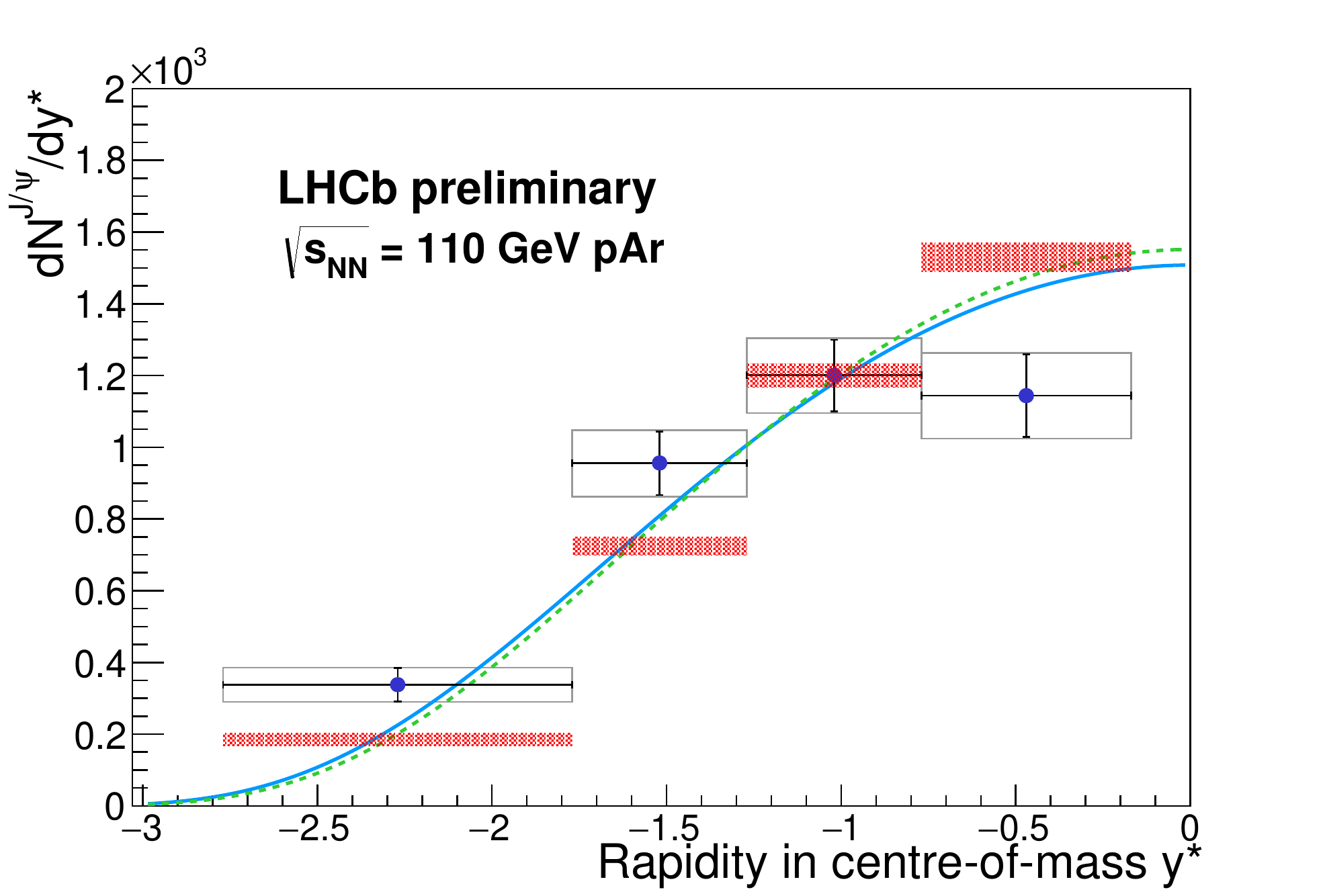}
 		\caption{$J/\psi$ differential corrected yields as a function of transverse momentum $p_T$ (left) and rapidity in the centre-of-mass system of the  			nucleon-nucleon collision $y^*$ (right). The blue and green curves are phenomenological parametrization based on~\cite{Arleo2013},  respectively obtained  with linear and 			logarithmic interpolations between the results obtained in pC collisions by the HERA-B experiment at lower energy.   See Fig.~\ref{fig2} for graphical conventions. \label{fig3}}
 	\end{center}
 \end{figure}

\begin{figure}
	\begin{center}
		\includegraphics[width=.4\textwidth]{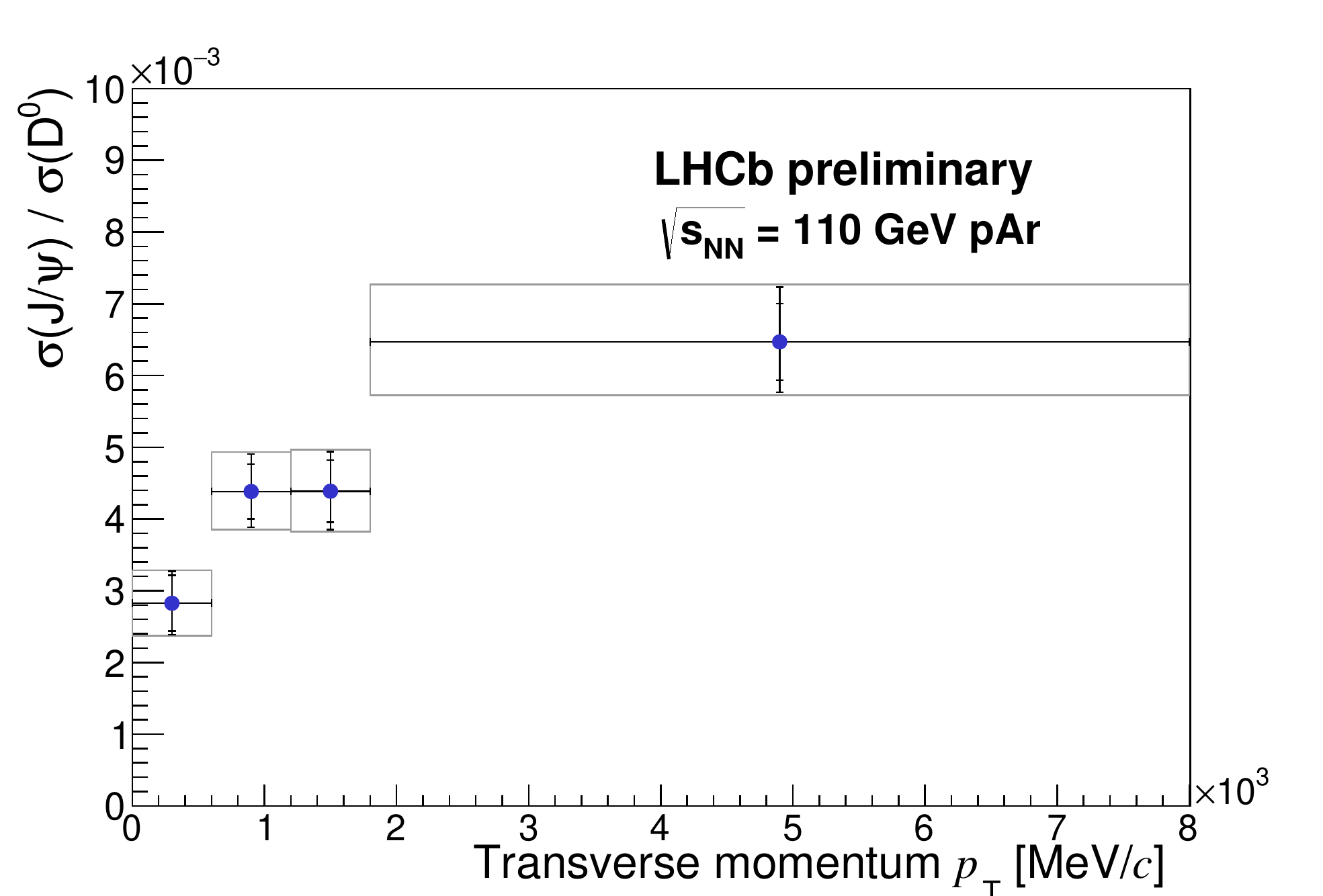}
		\includegraphics[width=.4\textwidth]{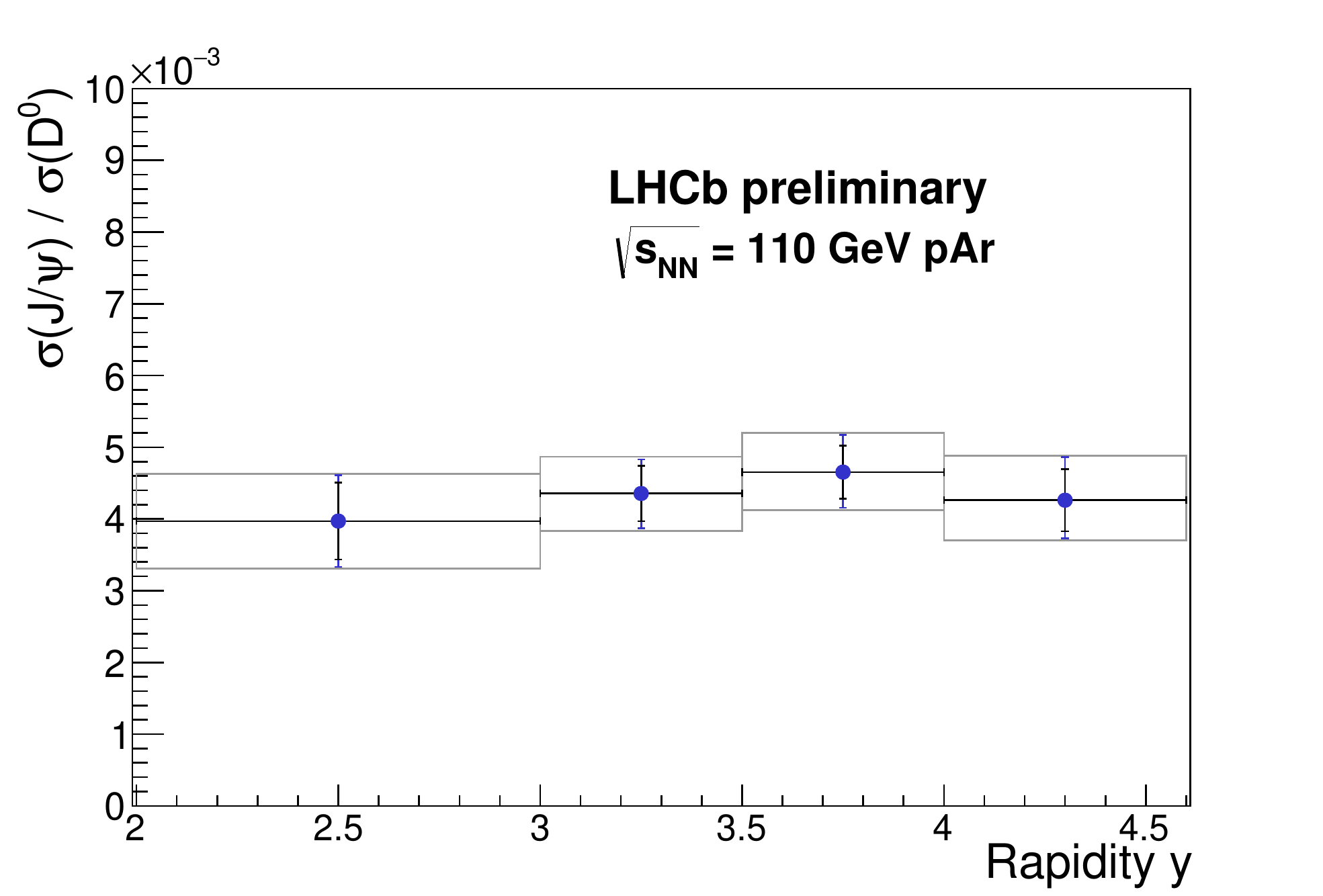}
		\caption{Ratio between $J/\psi$ and $D^0$ yields as a function of their transverse momentum, $p_T$ (left), and their rapidity, $y$ (right). 
			\label{fig_ratio}}
	\end{center}
\end{figure}

 Several systematic uncertainties affect either the determination of the number of signal events or the computation of the efficiencies. They are computed in every rapidity  and transverse momentum interval. The most important systematic uncertainties come from 
 the differences between $p$Ar and $pp$ track multiplicity for the efficiency computation, from the  contamination left from residual $pp$ collisions, from the signal and background modelling, and from the neglected non-prompt contribution of $J/\psi$ and $D^0$.  The fraction of signal from $b$-hadron is evaluated to $(0.9 \pm 1.6)\%$ .
 The differential $D^0$ yields corrected from the efficiency are shown in Fig~\ref{fig2}. Figure~\ref{fig3} presents  $J/\psi$ mesons corrected yields as a function of the  rapidity in the centre-of-mass system of the nucleon-nucleon collision, $y^{*}$ (given by: $y^{∗} = y - 4.77$, where 4.77 is the rapidity of the centre-of-mass in the LHCb frame at 110.4 GeV) and transverse momentum compared to phenomenological parametrizations based on Ref~\cite{Arleo2013}.
  The ratio between the  $J/\psi$ and $D^0$ yields as a function of their transverse momentum and their rapidity is shown in Fig~\ref{fig_ratio}.
 
\section{Conclusions}

The two first production measurements exploiting the LHCb experiment in fixed target mode have been presented. 

The antiproton production cross-section in collisions of a 6.5 TeV LHC proton beam on helium at rest is measured from a dataset corresponding to an integrated luminosity of 0.4 nb$^{-1}$. This is the first direct measurement of antimatter production in $p$He collisions and has important implication for the interpretation of recent results from PAMELA and AMS-02 experiments, which measure the antiproton component in cosmic rays outside of the Earth's atmosphere.

The production of $J/\psi$ and $D^0$ mesons is measured in 6.5 TeV$/ c$ proton-beam-induced reactions on an argon gaseous target, corresponding to about 500 selected $J/\psi$ and 6500 $D^0$ decays.  The $J/\psi$ and $D^0$ yields are computed as a function of their transverse momentum $p_T$ and rapidity $y$.  No significant difference,  within  uncertainties,  is  observed  when  comparing the $J/\psi$ and $D^0$ transverse momentum  shapes  obtained  with  the Pythia generator  in $pp$ collisions,  while  some differences are observed when comparing rapidity shapes.
The $J/\psi$ $/$ $D^0$ ratios  were  produced  as  a  function  of transverse momentum and rapidity.   Within  uncertainties, no strong dependence of this ratio with rapidity is observed.  This analysis demonstrates the feasibility of the heavy flavour fixed-target program. Future analysis should also include other open charm mesons or baryons such  as $\Lambda_c^+$.

\newpage

\Acknowledgements
The research leading to these results has received funding from the P2IO LabEx (ANR-10-LABX-0038) in the framework ``Investissements d'Avenir'' (ANR-11-IDEX-0003-01) managed by the Agence Nationale de la Recherche (ANR, France), from the People Programme (Marie Curie Actions) of the European Union’s Seventh Framework Programme (FP7/2007-2013) under REA grant agreement n. PCOFUND-GA-2013-609102, through the PRESTIGE programme coordinated by Campus France, and from the European Research Council (ERC) under the European Union's Horizon 2020 research and innovation programme (grant agreement n°ERC-2014-CoG-647390).

\end{document}

%% file: econfmacros.tex
\def\beq{\begin{equation}}
\def\eeq#1{\label{#1}\end{equation}}
\def\eeqn{\end{equation}}

\def\beqa{\begin{eqnarray}}
\def\eeqa#1{\label{#1}\end{eqnarray}}
\def\eeqan{\end{eqnarray}}

\let\bar=\overbar

\def\Dslash{\not{\hbox{\kern-4pt $D$}}}
\def\dslash{\not{\hbox{\kern-2pt $\del$}}}

\def\msb{{\bar{\ssstyle M \kern -1pt S}}}




%% file: eprint.bbl
\begin{thebibliography}{99}

\bibitem{LHCb}
LHCb collaboration, A. A. Alves Jr. et al., The LHCb detector at the LHC, JINST {\bfseries{3}} (2008) S08005

\bibitem{Lumi}

LHCb collaboration, R. Aaij et al., Precision luminosity measurements at LHCb, JINST {\bfseries{9}} (2014) P12005

\bibitem{Antiproton}

LHCb collaboration,  R. Aaij et al., Measurement of antiproton production in $p$He collisions at $\sqrt{s_{\scriptscriptstyle\rm NN}}=110$ GeV, LHCb-CONF-2017-002

\bibitem{EPOS}
	T. Pierog et al.,
	EPOS LHC: Test of collective hadronization with data measured at the CERN Large Hadron Collider,
	Phys. Rev. {\bfseries{C92}} (2015) 034906,


\bibitem{CRMC}
	T. Pierog et al., CRMS (Cosmic Ray Monte Carlo package),
	https://web.ikp.kit.edu/rulrich/crmc.html

\bibitem{EPOS_1.9.9}
	T. Pierog et al., 
	EPOS Model and Ultra High Energy Cosmic Rays,
	Nucl. Phys. Proc. Suppl. {\bfseries{196}} (2009) 102
		
\bibitem{HIJING}	
	M. Gyulassy and X.-N. Wang,
	HIJING 1.0: A Monte Carlo program for parton and particle production in high energy hadronic and nuclear collisions,
	Comput. Phys. Commun. {\bfseries 83} (1994) 307
		
	
\bibitem{QGSJET}
	S. Ostapchenko, 
  	Monte Carlo treatment of hadronic interactions in enhanced Pomeron scheme: QGSJET-II model,
  	Phys. Rev. {\bfseries D 83} (2011) 014018
  	
  
\bibitem{Chatrchyan:2012ufa} 
  S.~Chatrchyan {\it et al.}  [CMS Collaboration],
  Phys.\ Lett.\ B {\bf 716}, 30 (2012)

\bibitem{Satz_2006}
H. Satz, Colour deconfinement and quarkonium binding, J. Phys. \textbf{G32}(2006) R25

\bibitem{pArnote}
LHCb collaboration,  R. Aaij et al., Measurement of $J/\psi$ and $D^0$ production in $p$Ar collisions at $\sqrt{s_{NN}}=110$ GeV, LHCb-CONF-2017-001

\bibitem{LHCb-PUB-2016-021}
L. Anderlini et al., The PIDCalib package, LHCb-PUB-2016-021

\bibitem{Arleo2013}
F. Arleo et al., Centrality and p$_T$ dependence of $J/\psi$ suppression in proton-nucleus
collisions from parton energy loss,  JHEP \textbf{05} (2013) 155



\end{thebibliography}
